\RequirePackage{amsmath}
\documentclass[runningheads]{llncs}
\usepackage[utf8]{inputenc}
\usepackage[T1]{fontenc}
\usepackage{amsfonts}
\usepackage{mathtools}
\usepackage{amssymb}
\usepackage{color}
\usepackage{xcolor}

\newcommand{\bb}{\mathfrak{BB}}
\newcommand{\bigO}{\mathcal{O}}
\newcommand{\mc}[1]{\mathcal{#1}}

\usepackage{fancyhdr}
\title{Coercion-Resistant Voting in Linear Time via Fully Homomorphic Encryption}
\subtitle{Towards a Quantum-Safe Scheme}

\titlerunning{Coercion-Resistant Linear-Time Voting via Fully Homomorphic Encryption}

\author{Peter B. R\o nne\inst{1} \and Arash Atashpendar\inst{1}
\and Kristian Gj\o steen\inst{2}
\and \\ Peter Y. A. Ryan\inst{1}}

\authorrunning{Peter B. R\o nne \and Arash Atashpendar
\and Kristian Gj\o steen
\and Peter Y. A. Ryan}

\institute{SnT, University of Luxembourg, Luxembourg \email{\{peter.roenne,arash.atashpendar.peter.ryan\}@uni.lu}
\and Norwegian University of Science and Technology, NTNU, Norway \email{kristian.gjosteen@ntnu.no}
}

\begin{document}

\maketitle
\thispagestyle{fancy} %This is just to make the preprint statement appear at the bottom of page 1.

\begin{abstract}
We present an approach for performing the tallying work in the coercion-resistant JCJ voting protocol, introduced by Juels, Catalano, and Jakobsson, in linear time using fully homomorphic encryption (FHE). The suggested enhancement also paves the path towards making JCJ quantum-resistant, while leaving the underlying structure of JCJ intact. The pairwise comparison-based approach of JCJ using plaintext equivalence tests leads to a quadratic blow-up in the number of votes, which makes the tallying process rather impractical in realistic settings with a large number of voters. We show how the removal of invalid votes can be done in linear time via a solution based on recent advances in various FHE primitives such as hashing, zero-knowledge proofs of correct decryption, verifiable shuffles and threshold FHE. We conclude by touching upon some of the advantages and challenges of such an approach, followed by a discussion of further security and post-quantum considerations.
\end{abstract}

\section{Introduction}

Over the past few decades, we have witnessed significant advances in cryptographic voting protocols. Yet, despite all the progress, see e.g., \cite{adida2008helios}, secure e-voting is still faced with a plethora of challenges and open questions, which largely arise as a result of the interplay between intricate properties such as vote privacy, individual and universal verifiability, receipt-freeness, and a notoriously difficult requirement, namely that of coercion-resistance. Coercion-resistance can be viewed as a stronger form of privacy that should hold even against an adversary who may instruct honest parties to carry out certain computations while potentially even requiring that they reveal secrets in order to verify their behavior and ensure compliance. This property is typically enforced by providing honest parties with a mechanism that allows them to either deceive the coercer or to deny having performed a particular action. Due to limited space, we do not elaborate on the long series of works in this area and instead refer the reader to \cite{juels2005coercion,delaune2006coercion,kusters2012game,cortier2016sok}
and references therein for more details.

Since the breakthrough work of Gentry \cite{gentry2009fully} on fully homomorphic encryption (FHE), there has been a surge of interest in this line of research that remains very active to this day, with a series of recent advances including, but not limited to, a homomorphic evaluation of AES \cite{gentry2012homomorphic}. Although the use of additively or multiplicatively homomorphic cryptosystems is common place in the e-voting literature, the relevance of FHE for potentially quantum-safe secure e-voting, with better voter verifiability, was only recently discussed by Gj\o steen and Strand \cite{gjosteen2017roadmap}. In our work, instead of designing an FHE-based protocol from scratch, we apply the machinery of FHE to a well-known, classical voting scheme, in order to improve its time complexity and to replace its reliance on the hardness assumption of solving the discrete logarithm problem with a quantum-resistant solution, namely lattice-based cryptography. So far, no efficient quantum algorithms capable of breaking lattice-based FHE schemes have been discovered.

\fancyhead[LO]{Coercion-Resistant Voting in Linear Time via Fully Homomorphic Encryption}
\fancyhead[RE]{Peter B. R\o nne, Arash Atashpendar, Kristian Gj\o steen, Peter Y. A. Ryan}

Although constructions with varying degrees of coercion-resistance do exist, the voting protocol introduced by Juels, Catalano, and Jakobsson \cite{juels2005coercion}, often referred to as the \emph{JCJ protocol}, is among the most well-known solutions in the context of coercion-resistant voting schemes. JCJ provides a reasonable level of coercion-resistance using a voter credential faking mechanism, and it was arguably the first proposal with a formal definition of coercion-resistance. However, JCJ suffers from a complexity problem due to the weeding steps in its tallying phase, which are required for eliminating invalid votes and duplicates. The exhaustive, comparison-based approach of JCJ using plaintext equivalence tests (PET) \cite{jakobsson2000mix}  leads to a quadratic blow-up in the number of votes, which makes the tallying process rather impractical in realistic settings with a large number of voters or in the face of ballot-box stuffing. For instance, in the Civitas voting system \cite{DBLP:conf/sp/ClarksonCM08} based on JCJ, voters are grouped into blocks or virtual precincts to reduce the tallying time.

Here we propose an enhancement of the JCJ protocol aimed at performing its tallying work in linear time, based on an approach that incorporates primitives from the realm of fully homomorphic encryption (FHE), which also paves the path towards making JCJ quantum-safe.

In Sect. \ref{sec:jcj}, we describe the JCJ protocol and cover some related work. Next, in Sect. \ref{sec:jcj-with-fhe}, we show how the weeding of ``bad'' votes can be done in linear time, with minimal change to JCJ, via an approach based on recent advances in various FHE primitives such as hashing, zero-knowledge (ZK) proofs of correct decryption, verifiable shuffles and threshold FHE. We also touch upon some of the advantages and remaining challenges of such an approach in Sect. \ref{subsec:open-questions} and in Sect. \ref{sec:security-considerations}, we discuss further security and post-quantum considerations.

\section{The JCJ Model and Voting Protocol in a Nutshell}\label{sec:jcj}

\paragraph{\textbf{Cryptographic Building Blocks.}}

JCJ relies on a modified version of ElGamal, a threshold public-key cryptosystem with re-encryption, secure under the hardness assumption of the Decisional Diffie-Hellman (DDH) problem in a multiplicative cyclic group $\mc{G}$ of order $q$. A ciphertext on message $m \in \mc{G}$ has the form $(\alpha, \beta, \gamma) = (mh^r, g_1^r, g_2^r)$ for $r \in_U \mathbb{Z}_q$, with $(g_1, g_2, h)$ being the public key where $g_1,g_2,h \in \mc{G}$, and the secret key consists of $x_1,x_2 \in \mathbb{Z}_q$ such that $h=g_1^{x_1}g_2^{x_2}$. The construction allows easy sharing of the secret key in a threshold way.
The weeding steps make use of a plaintext equivalence test (PET), which is carried out by the secret key holders and takes as input two ciphertexts and outputs 1 if the underlying plaintexts are equal, and 0 otherwise. The PET produces publicly verifiable evidence with negligible information leakage about plaintexts.  Finally, JCJ uses non-interactive zero-knowledge (NIZK) proofs and mix-nets, which are aimed at randomly and secretly permuting and re-encrypting input ciphertexts such that output ciphertexts cannot be traced back to their corresponding ciphertexts. Throughout, it is assumed that the computations of the talliers and registrars are done in a joint, distributed threshold manner. We use $\in_U$ to denote an element that is sampled uniformly at random.

\paragraph{\textbf{Agents.}}

JCJ mainly consists of three sets of agents, described as follows.
\begin{enumerate}
	\item \textbf{Registrars}: A set $\mathcal{R} = \{R_1, R_2, \ldots, R_{n_R}\}$ of $n_R$ entities in charge of jointly generating and distributing credentials to voters.

	\item \textbf{Talliers}: A set $\mathcal{T} = \{T_1, T_2, \ldots, T_{n_T} \}$ of \emph{authorities} in charge of processing ballots, jointly counting the votes and publishing the final tally.

	\item \textbf{Voters}: A set of $n_V$ voters, $\mathcal{V} = \{V_1, V_2, \ldots, V_{n_V} \}$, participating in an election, where each voter $V_i$ is publicly identified by an index $i$.
\end{enumerate}

\paragraph{\textbf{Bulletin Board and Candidate Slate.}}

A \emph{bulletin board}, denoted by $\bb$, is an abstraction representing a publicly accessible, append-only, but otherwise immutable board, meaning that participants can only add entries to $\bb$ without overwriting or erasing existing items. A \emph{candidate slate}, $\vec{C}$, is an ordered set of $n_C$ distinct identifiers $\{c_1, c_2, \ldots, c_{n_C}\}$ capturing voter choices.
A $\emph{tally}$ is defined under slate $\vec{C}$, as a vector $\vec{X} = \{x_1, x_2, \ldots, x_{n_C}\}$ of $n_C$ positive integers, where each $x_j$ indicates the number of votes cast for choice $c_j$.

\paragraph{\textbf{Assumptions for Coercion-Resistance.}}
No threshold set of agents in $\mc{T}$ should be corrupted, otherwise privacy is lost. In the registration phase, it is assumed that the distribution of voter credentials is done over an untappable channel and that no registration transcripts can be obtained, assuming that secure erasure is possible. Cast votes are transmitted via anonymous channels, which is a basic requirement for ruling out forced-abstention attacks.

\subsection{The JCJ Protocol}

\paragraph{\textbf{Setup and Registration.}}
The key pairs $(sk_{\mc{R}}, pk_{\mc{R}})$ and $(sk_{\mc{T}}, pk_{\mc{T}})$ are generated in a trustworthy manner, and the public keys, i.e., $pk_{\mc{T}}$ and $pk_{\mc{R}}$, are published with other public system parameters. The registrars $\mc{R}$ generate and transmit to eligible voter $V_i$ a random string $\sigma_i \in_U \mc{G}$ that serves as the credential of the voter. $\mc{R}$ adds an encryption of $\sigma_i$, $S_i = E_{pk_{\mc{T}}}(\sigma_i)$, to the voter roll $\vec{L}$, which is maintained on the bulletin board $\bb$ and digitally signed by $\mc{R}$.

\paragraph{\textbf{Voting.}} An integrity-protected candidate slate $\vec{C}$ containing the names and unique identifiers in $\mathcal{G}$ for $n_C$ candidates, along with a unique, random election identifier $\epsilon$ are published by the authorities. Voter $V_i$ generates a ballot in the form of a variant of ElGamal ciphertexts $(E_1, E_2)$, for candidate choice $c_j$ and voter credential $\sigma_i$, respectively, e.g., for $a_1, a_2 \in_U \mathbb{Z}_q$, we have $E_1 = (g_1^{a_1}, g_2^{a_1}, c_j h^{a_1})$ and $E_2 = (g_1^{a_2}, g_2^{a_2}, \sigma_i h^{a_2})$.
$V_i$ computes NIZK proofs of knowledge and correctness of $\sigma_i$ and $c_j \in \vec{C}$, collectively denoted by $P_f$. These ensure non-malleability of ballots, also across elections by including $\epsilon$ in the hash of the Fiar-Shamir heuristic. $V_i$ posts $B_i = (E_1, E_2, P_f)$ to $\bb$ via an anonymous channel.

\paragraph{\textbf{Tallying.}} In order to compute the tally, duplicate votes and those with invalid credentials will have to be removed. The complexity problem crops up in steps \colorbox{lightgray}{2} and \colorbox{lightgray}{4} %of the tallying phase
such that given $n$ votes, the tallying work has a time complexity of $\bigO(n^2)$. To tally the ballots posted to $\bb$, the authority $\mc{T}$ performs the following steps:
\colorbox{lightgray}{1.} $\mc{T}$ verifies all proofs on $\bb$ and discards any ballots with invalid proofs.
Let $\vec{A_1}$ and $\vec{B_1}$ denote the list of remaining $E_1$ candidate choice ciphertexts, and $E_2$ credential ciphertexts, respectively.
\colorbox{lightgray}{2.} $\mc{T}$ performs pairwise PETs on all ciphertexts in $\vec{B_1}$ and removes duplicates according to some fixed criterion such as the order of postings to $\bb$. For every element removed from $\vec{B_1}$, the corresponding element with the same index is also removed from $\vec{A_1}$, resulting in the ``weeded'' vectors $\vec{B'_1}$ and $\vec{A'_1}$.
\colorbox{lightgray}{3.} $\mc{T}$ applies a mix-net to $\vec{A'_1}$ and $\vec{B'_1}$ using the same, secret permutation, resulting in the lists of ciphertexts $\vec{A_2}$ and $\vec{B_2}$.
\colorbox{lightgray}{4.} $\mc{T}$ applies a mix-net to the encrypted list $\vec{L}$ of credentials from the voter roll and then compares each ciphertext of $\vec{B_2}$ to the ciphertexts of $\vec{L}$ using a PET. $\mathcal{T}$ keeps a vector $\vec{A_3}$ of all ciphertexts of $\vec{A_2}$ for which the corresponding elements of $\vec{B_2}$ match an element of $\vec{L}$, thus achieving the weeding of ballots with invalid voter credentials.
\colorbox{lightgray}{5.} $\mc{T}$ decrypts all ciphertexts in $\mathbf{A_3}$ and tallies the final result.

\paragraph{Properties.}

Vote \textbf{privacy} is maintained as long as neither a threshold set of talliers nor all the mixing servers are corrupted. A colluding majority of talliers can obviously decrypt everything and colluding mixing authorities could trace votes back to $\vec{L}$. Regarding \textbf{correctness}, voters can refer to $\bb$ to verify that their vote has been recorded as intended and that the tally is computed correctly. Similar attacks become possible in case of collusion by a majority of authorities. As for \textbf{verifiability}, anyone can refer to $\bb$, $P_f$ and $\vec{L}$ to verify the correctness of the tally produced by $\mc{T}$.
The \textbf{coercion-resistance} provided by JCJ essentially comes from %boils down to
keeping voter credentials hidden throughout the election. A coerced voter can then choose a random fake credential $\sigma'$ to cast a fake vote and present it as their real vote. Any vote cast with the fake credential will not be counted, and the voter can anonymously cast their real vote using their real credential.

\subsection{Related Work}\label{subsec:related-work}

We focus on the most closely-related works on improving the efficiency problem of the tallying work in JCJ. Smith \cite{smith2005new} and Weber et al. \cite{weber2007coercion,weber2008coercion} follow a similar approach in that they do away with comparisons using PETs, and instead, they raise the credentials to a jointly $\mc{T}$-shared secret value and store these blinded terms in a hash table such that collisions can be found in linear time.
The use of a single exponent means that a coercer can test if the voter has provided them with a fake or a real credential by submitting a ballot with the given credential and another with the credential raised to a known random value.
In \cite{araujo2008practical,araujo2010towards}, Araujo et al. move away from comparing entries in $\vec{L}$ with terms in the cast ballots to a setting in which duplicates are publicly identifiable and a majority of talliers use their private keys to identify legitimate votes, and in \cite{DBLP:conf/fc/AraujoBBT16} the authors use algebraic MACs. Spycher et al. \cite{spycher2011new} use the same solution proposed by Smith and Weber to remove duplicates and apply targeted PETs only to terms in $\vec{L}$ and $\vec{A}$, identified via additional information provided by voters linking their vote to the right entry in $\vec{L}$. In \cite{grontas2018towards}, publicly auditable conditional blind signatures are used to achieve coercion-resistance %and everlasting privacy
in linear time using a FOO-like \cite{fujioka1992practical} architecture, the downsides being the need for extra authorization requests for participation privacy and a double use of anonymous channels.

\section{JCJ in Linear Time via Fully Homomorphic Encryption}\label{sec:jcj-with-fhe}

Our proposal revolves around replacing the original cryptosystem of JCJ with a fully homomorphic one, thus allowing us to preserve the original design of JCJ. The main idea is to homomorphically evaluate hashes of the underlying plaintext of the FHE-encrypted voter credentials, perform FHE-decryption and post the hash values of the credentials to the bulletin board $\bb$. Now the elimination of invalid and duplicate entries can be done in linear time by using a hash table.

\paragraph{\textbf{FHE Primitives.}}

Constrained by limited space, we only enumerate the cryptographic primitives that will be required for the enhancement suggested below and refer the reader to the cited sources for further details. Let $\mc{E}_{pk}(m)$ denote an FHE-encryption of a message $m \in \{0,1\}^n$ under the public key $pk$. At its core, for $b_0, b_1 \in \{0,1\}$, given $\mc{E}_{pk}(b_0)$ and $\mc{E}_{pk}(b_1)$, FHE allows us to compute $\mc{E}_{pk}(b_0 \oplus b_1)$ and $\mc{E}_{pk}(b_0 \cdot b_1)$ by working over ciphertexts alone, without having access to the secret key, thus enabling the homomorphic evaluation of any boolean circuit, i.e., computing $\mc{E}_{pk}(f(m))$ from $\mc{E}_{pk}(m)$ for any computable function $f$. We make use of FHE \cite{gentry2009fully,brakerski2014leveled}, fully homomorphic hashing \cite{fiore2014efficiently}, zero-knowledge proofs of correct decryption for FHE ciphertexts \cite{carr2018zero}, verifiable shuffles \cite{strand2018fhemixnet} and threshold FHE \cite{boneh2018threshold}, see Sect. \ref{subsec:open-questions} for more details on open questions and the state-of-the-art.

\subsection{Enhancing JCJ with FHE and Weeding in Linear Time}

We now describe how FHE primitives can be incorporated into JCJ while inducing minimal change in the original protocol. We assume threshold FHE throughout.

\paragraph{\textbf{Setup and Registration.}}

The setup and registration phases remain unchanged w.r.t. JCJ, except that $\mc{R}$ now adds an FHE-encryption of $\sigma_i$, $S_i = \mc{E}_{pk_{\mc{T}}}(\sigma_i)$, to the voter roll $\vec{L}$. We adopt the same assumptions mentioned earlier in Sect. \ref{sec:jcj}.

\paragraph{\textbf{Voting.}}

Instead of using ElGamal encryption, the credentials posted on the $\bb$ are encrypted under some FHE scheme, say BGV \cite{brakerski2014leveled}, with a key pair $(pk,sk)$. Each voter $V_i$ adds $\mc{E}_{pk_{\mc{T}}}(\sigma_i)$, along with the required NIZK proofs, to $\bb$.

\paragraph{\textbf{Tallying.}}

The tallying phase remains largely the same except that for removing duplicates and invalid votes, we leverage our use of FHE to carry out simple equality tests between hash digests of credentials. Since the concealed credentials are now stored in FHE ciphertexts, we can process them using an FHE hashing circuit. More precisely, for a jointly created $\mc{T}$-shared key $k$, published under encryption $\mc{E}_{pk}(k)$, the credentials $\sigma_i$ contained in the FHE-encrypted terms $\mc{E}_{pk}(\sigma_i)$ are homomorphically hashed (see \cite{fiore2014efficiently} by Fiore, Gennaro and Pastro and \cite{catalano2014authenticating} by Catalano et al.), under key $k$ resulting in $\mc{E}_{pk}(h_k(\sigma_i))$, such that upon decryption we obtain $h_k(\sigma_i)$.
A ZK proof of correct decryption is also posted to $\bb$ for verifiability, see \cite{carr2018zero} by Carr et al. for an approach to this. Once the hash values of the credentials are posted on the $\bb$, the weeding of duplicates can be done in $\bigO(n)$ using a simple hash table look-up, i.e., iterate, hash and check for collision in constant time, thus an overall linear-time complexity in the number of votes. Next, the registered credentials and the submitted vote/credential pairs are mixed \cite{strand2018fhemixnet} and the homomorphic hashing procedure is carried out again using a new secret key on all credential ciphertexts. Comparing the hashed registered credentials with those from the cast ballots allows us to remove invalid votes in $\bigO(n)$. Finally, the remaining valid votes are verifiably decrypted.

\subsection{Advantages, Potential Pitfalls and Open Questions}\label{subsec:open-questions}

Apart from the linear-time weeding algorithm, as already pointed out by Gj\o steen and Strand in \cite{gjosteen2017roadmap}, in addition to being a novel application of FHE to secure e-voting, obtaining better voter verifiability and a scheme believed to be quantum-resistant are among the noteworthy benefits of such an approach.
Clearly, in terms of real world FHE implementations, the state-of-the-art still suffers from efficiency issues. However, some significant progress has already been made in this area, e.g., the homomorphic evaluation of AES \cite{gentry2012homomorphic} or block ciphers designed for homomorphic evaluation \cite{DBLP:conf/eurocrypt/AlbrechtR0TZ15}. Moreover, it should be pointed out that some of the needed primitives, e.g., turning ZK proofs of correct decryption for FHE \cite{carr2018zero,luo2018verifiable} into NIZK proofs, are still not satisfactory and remain the subject of ongoing research and future improvements.

\section{Further Security Remarks}\label{sec:security-considerations}

A security analysis aimed at providing proofs of security for various properties such as correctness, verifiability and coercion-resistance will remain future work. One possibility would be to investigate whether the required security properties in our enhanced variant of JCJ hold against classical adversaries, under the same oracle access assumptions for mixing, PETs, threshold decryption and hashing. Post-quantum security will have to be proved in the quantum random oracle model.

\paragraph{\textbf{Eligibility Verifiability.}}

Assuming a majority of colluding authorities, apart from a compromise of vote privacy, another, perhaps more damaging problem with JCJ and its improved variants is that of \emph{eligibility verifiability}. A colluding majority would be able to retrieve voter credentials and submit valid votes for non-participating voters, i.e., perform ballot stuffing.
A solution in \cite{RoenneJCJ16} suggests performing the registration phase in such a way that only the voter would know the discrete logarithm of their credential. Votes are then cast with an anonymous signature in the form of a ZK proof of knowledge of the discrete logarithm of the encrypted credential, thus preventing ballot stuffing. A similar approach could be used here, with the potential downside of having inefficient proofs and a discrete logarithm hardness assumption, thus not being quantum secure.

\paragraph{\textbf{Post-Quantum Considerations.}}

For a relaxation of the trustworthiness assumption of $\mc{R}$, without assuming secure erasure, quantum-resistant designated verifier proofs \cite{sun2012toward,jao2014isogeny} could replace the classical ones suggested in the original JCJ \cite{juels2005coercion}.
To obtain post-quantum security for eligibility verifiability, future research will investigate the use of a quantum-resistant signature scheme that can be evaluated under FHE to preserve ballot anonymity. As a naive, but illustrative example that is one-time only and non-distributive, consider that the voter creates their credential as $\sigma_i = h(x)$, and that only the voter knows the preimage $x$. The voter now submits both $\mc{E}_{pk}(x)$ and $\mc{E}_{pk}(\sigma_i)$ to $\bb$. Before weeding, the hash is homomorphically evaluated on the ciphertext of the preimage, i.e., $\mc{E}_{pk}(h(x))$, followed by an equality test against the ciphertext of the credential $\mc{E}_{pk}(\sigma_i)$. A malicious authority can now cast only a valid ballot with a registered credential after the corresponding voter has cast a ballot, and an attempt to vote on their behalf is detectable in the weeding phase.

\subsection*{Acknowledgments}

The authors acknowledge support from the Luxembourg National Research Fund (FNR) and the Research Council of Norway for the joint project SURCVS. The project was also supported by the FNR INTER-VoteVerif and the FNR CORE project Q-CoDe.

\bibliographystyle{splncs04}
\bibliography{LinJCJ}

\end{document}